\documentclass[runningheads]{llncs}
\usepackage{graphicx}
\usepackage{textcomp}
\usepackage{comment}

\usepackage{color}
\usepackage{graphicx}
\usepackage{amsmath}
\usepackage{amssymb}
\usepackage{booktabs}
\usepackage{verbatim}
\usepackage{import}
\usepackage{multirow}
\usepackage{hyperref}
\usepackage[capitalize]{cleveref}
\usepackage{caption}
\usepackage{subcaption}
\usepackage[utf8]{inputenc}
\usepackage[english]{babel}
\usepackage{biblatex}
\usepackage{csquotes}
\usepackage{adjustbox}
\usepackage{float}
\restylefloat{table}
\usepackage{placeins}
\usepackage{xspace}
\makeatletter
\DeclareRobustCommand\onedot{\futurelet\@let@token\@onedot}
\def\@onedot{\ifx\@let@token.\else.\null\fi\xspace}

\def\ie{\emph{i.e}\onedot} 
 
 \def\vs{\emph{vs}\onedot}

\makeatother

\newcommand*{\figuretitle}[1]{{\centering \textbf{#1}\par\medskip}}
\usepackage{todonotes}

\usepackage{float}
\floatstyle{plaintop}
\restylefloat{table}

\begin{document}
\titlerunning{Impact of model choice on bias in cardiac MR segmentation}

\title{An investigation into the impact of deep learning model choice on sex and race bias in cardiac MR segmentation}

\newcommand*\samethanks[1][\value{footnote}]{\footnotemark[#1]}
\newcommand{\rowstyle}[1]{\gdef\currentrowstyle{#1}%
  #1\ignorespaces
}

\author{Tiarna Lee \inst{1}
\and Esther Puyol-Ant\'on \inst{1, 4} \and 
Bram Ruijsink \inst{1,2}
\and Keana Aitcheson \inst{1}
\and Miaojing Shi \inst{3}
\and Andrew P. King \inst{1} }
\authorrunning{T Lee et al.}   
\institute{School of Biomedical Engineering \& Imaging Sciences, King\textquotesingle s College London, UK. \and St Thomas\textquotesingle{} Guy’s and St Thomas’ Hospital, London, UK. \and College of Electronic and Information Engineering, Tongji University, China. \and HeartFlow Inc, London, UK.}
\maketitle              

\begin{abstract} 
In medical imaging, artificial intelligence (AI) is increasingly being used to automate routine tasks. However, these algorithms can exhibit and exacerbate biases which lead to disparate performances between protected groups. We investigate the impact of model choice on how imbalances in subject sex and race in training datasets affect AI-based cine cardiac magnetic resonance image segmentation. We evaluate three convolutional neural network-based models and one vision transformer model. We find significant sex bias in three of the four models and racial bias in all of the models. However, the severity and nature of the bias varies between the models, highlighting the importance of model choice when attempting to train fair AI-based segmentation models for medical imaging tasks. 

\end{abstract}

\section{Introduction}
\label{sec:intro}

The popularity and increasing clinical translation of artificial intelligence (AI) techniques in medical image analysis have prompted many to investigate their fairness. Fairness can be defined in a number of ways \cite{Mehrabi2019ALearning} but in simple terms, it refers to the absence of inexplicable bias, or disparate performance, between different protected groups. Biases, which cause performances to be unfair, can have several causes. One such cause is training models using datasets that are imbalanced by protected group and recent work has shown that this can lead to bias in a range of medical imaging applications \cite{Larrazabal2020,Seyyed-Kalantari2021a,Puyol-Anton2021FairnessSegmentation, Lee2022ASegmentation, Petersen2022FeatureDetection}. For example, \cite{Larrazabal2020} found that accuracy for a chest X-ray classification task was significantly lower for a protected group, in this case sex, that was under-represented in the training dataset.

Although under-representation in the training data can be a root cause of bias in AI, there has been relatively little research into the impact of other design choices on bias. One such choice is model architecture. To date, we are aware of no work in medical image analysis that has compared biases between different models trained for the same task with the same data. Furthermore, we are aware of no work that has aimed to assess the potential for bias in the latest vision transformer based architectures in medical image analysis. Therefore, in this work we aim to perform a comparative investigation of the impact of model architecture on bias, including both convolutional neural network (CNN)-based models and vision transformers.

We investigate the potential for race and sex bias in deep learning-based cardiac magnetic resonance (CMR) segmentation, an area where bias has previously been reported \cite{Puyol-Anton2021FairnessSegmentation,Puyol-Anton2022FairnessSegmentation,Lee2022ASegmentation} for a CNN-based model (nnU-Net \cite{Isensee2020}). We perform a systematic analysis of the impact of training set imbalance on segmentation performance, similar to \cite{Lee2022ASegmentation}. However, we perform these experiments for four different deep learning-based architectures, including one based on vision transformers, and analyse their differences, both in terms of overall performance and bias.

To summarise, our key contributions are:

\begin{enumerate}
    \item We perform the first investigation into the impact of model architecture on AI model bias in medical image analysis.
    \item We perform the first investigation into potential bias in vision transformer-based models in medical image analysis.
\end{enumerate}

\section{Materials}
\label{sec:materials}

In this work, we used CMR images from the UK Biobank \cite{Peterson2016}. The dataset consists of end diastolic (ED) and end systolic (ES) cine short-axis images from 661 subjects. The demographic data for these subjects were also gathered from the UK Biobank database and can be seen in \cref{tab:subject demographics} for the subjects used. 

\begin{table*}[htpb]
  \caption{Clinical characteristics of subjects used in the experiments. Mean values are presented for each characteristic with standard deviations given in brackets. Statistically significant differences are indicated with an asterisk * (\(p<0.05\)) and were determined using a two-tailed Student's t-test.}
  \centering
  \resizebox{\textwidth}{!}{
  \begin{tabular}{c|c|c|c|c|c}
    \toprule
    Health measure & Overall &  Male & Female & White & Black \\
     \midrule
     \#Subjects & 661 & 321 & 340 & 427 & 234  \\
    Age (years) & 60.9 (8.1) & 61.0 (8.2) & 60.9 (8.0) & 62.1 (8.4)* & 58.8 (6.9) *  \\
    Weight (kg) & 79.3 (16.1) & 86.0 (14.9) * & 72.9 (14.4) * & 78.0 (16.0) & 81.6 (16.0) \\
    Standing height (cm) & 169.8 (9.5) & 176.7 (6.6) * & 163.2 (6.9) * & 170.3 (9.6) & 168.8 (9.3)  \\
    Body Mass Index (kg) & 27.4 (4.7) & 27.5 (4.2) & 27.4 (5.1) & 26.8 (4.4)* & 28.6 (5.1) * \\
    \bottomrule
  \end{tabular}}

  \label{tab:subject demographics}
\end{table*}


Manual segmentation of the left ventricular blood pool (LVBP), left ventricular myocardium (LVM), and right ventricular blood pool (RVBP) was performed for the ED and ES images of each subject. This was done by outlining the LV endocardial and epicardial borders and the RV endocardial border using cvi42 (version 5.1.1, Circle Cardiovascular Imaging Inc., Calgary, Alberta, Canada). A panel of ten experts was provided with the same guidelines and one expert annotated each ground truth image. The selection of images for annotation was randomised and included subjects with different sexes and races. The experts were not provided with demographic information about the subjects such as their race or sex. 
\section{Methods}
\label{sec:methods}

\subsection{Dataset sampling}
We investigated the effect of varying the proportions of training subjects based on two protected attributes: male \vs female (Experiment 1) and White \vs Black (Experiment 2). In Experiment 1, the race of the subjects was controlled so that 50\% of male and female subjects were Black and 50\% were White. Similarly, in Experiment 2, the sex of the subjects was controlled so that 50\% of subjects from each race were female and 50\% were male. For each experiment, a group of 176 subjects were chosen for training from each protected group and the combination of these two groups was sampled to create five training sets which varied the proportion of subjects with the selected protected attribute from 0\%/100\% to 100\%/0\% as in \cite{Lee2022ASegmentation}. The total number of subjects in each of the five training sets was constant at 176. 

One difference in our experimental setup compared to \cite{Lee2022ASegmentation} was that, to remove the effect of a potential confounder in our analysis, the age of the subjects was controlled so that each subject in a protected group was matched to a subject in the other protected group whose age was within $\pm 1$ year. For both of the experiments, the test sets comprised 84 subjects and contained 50\% Black and 50\% White subjects, and 50\% male and 50\% female subjects. These subjects were also controlled for age.

\subsection{Model architecture and implementation}
All experiments were performed using four separate segmentation models: U-Net \cite{Ronneberger2015U-net:Segmentation}, nnU-Net \cite{Isensee2020}, Swin-Unet \cite{Cao2023Swin-Unet:Segmentation} and DeepLabv3+ \cite{Chen2018Encoder-decoderSegmentation}. For nnU-Net, we used the same training parameters as in \cite{Lee2022ASegmentation}.
Swin-Unet is a U-Net-like transformer model which uses an encoder-decoder structure based on Swin Transformer blocks and skip connections to merge the multi-scale features \cite{Cao2023Swin-Unet:Segmentation}. Training hyperparameters from \cite{Cao2023Swin-Unet:Segmentation} were used, which had been tuned using cine short axis CMR images from the ACDC dataset \cite{Bernard2018DeepSolved}. The model was pre-trained using data from ImageNet \cite{Deng2010ImageNet:Database}. The U-Net model was trained using the U-Net64 model from \cite{Chen2020ImprovingImages} and used the hyperparameters from their work which were chosen using CMR data from the UK Biobank. The final model was DeepLabv3+, an encoder-decoder architecture which utilises atrous spatial pyramid pooling and atrous convolutions \cite{Chen2018Encoder-decoderSegmentation}. The model was trained using a ResNet-50 backbone and hyperparameters were those reported in \cite{Chen2018Encoder-decoderSegmentation} which were chosen using the PASCAL VOC dataset \cite{Everingham2015TheRetrospective}. 

A summary of the training setup for each of the four models is provided in \cref{tab:training_params}.
The same data augmentation methods were used for each of the models and are as described in \cite{Lee2022ASegmentation}. For each of the models, the images were cropped to 224 $\times$ 224. All models were optimised using Stochastic Gradient Descent. The models were implemented using Python and PyTorch, and were trained on one NVIDIA A100 GPU.

\subsection{Model evaluation}
\label{sec:method model evaluation}
Model performance was assessed using the \emph{Dice similarity coefficient} (DSC), which measures the spatial overlap between two sets. For a ground truth segmentation A and its corresponding prediction B, the DSC is given by ${DSC} = \frac{2|A\cap B|}{|A| +|B|}$.

To compare the fairness of the different models, the median DSC values for each protected group in the test set were first calculated. We denote these median DSC values by $\mathbf{ {D}}_{A_1} = [D_{A_1}^0, D_{A_1}^{25}, ... D_{A_1}^{100}]$ and $\mathbf{ {D}}_{A_2} = [D_{A_2}^{100}, D_{A_2}^{75}, ... D_{A_2}^{0}]$ where $A_1$ and $A_2$ are the protected groups and the superscripts indicate the percentage of the protected group in the training set. 
Based upon $\mathbf{ {D}}_{A_1}$ and $\mathbf{ {D}}_{A_2}$, the following metrics were calculated:
\begin{itemize}
    \item \textbf{Fairness gap}: 
    For a given model and level of imbalance, the fairness gap is found by subtracting the median DSCs of the model evaluated on the two protected groups. We report the maximum and minimum fairness gap across all levels of imbalance, i.e. $FG_{max} = \max(\mathbf{ {D}}_{A_1} - \mathbf{ {D}}_{A_2})$ and $FG_{min} = \min(\mathbf{ {D}}_{A_1} - \mathbf{ {D}}_{A_2})$.
    These metrics quantify the level of performance disparity \textit{between} protected groups in both directions, i.e. in favour of $A_1$ (the maximum fairness gap) and in favour of $A_2$ (the minimum fairness gap).
    \item \textbf{Performance range}: The performance range for a given protected group is found by calculating the difference between the minimum and maximum median DSC scores across all levels of imbalance, i.e. $PR_{A_1} = \max(\mathbf{ {D}}_{A_1}) - \min(\mathbf{ {D}}_{A_1})$ and $PR_{A_2} = \max(\mathbf{ {D}}_{A_2}) - \min(\mathbf{ {D}}_{A_2})$. These metrics quantify differences in performance \textit{within} protected groups.
    \item \textbf{Skewed error ratio}: For protected groups $A_1$ and $A_2$, the skewed error ratio is defined by $SER_{A_1} = \frac{\max (1 -\mathbf{ {D}_{A_1}})}{\min (1- \mathbf{ {D}_{A_1}})}$ and $SER_{A_2} = \frac{\max (1 -\mathbf{ {D}_{A_2}})}{\min (1- \mathbf{ {D}_{A_2}})}$. This represents the ratio between the error rate of the best performing model and the worst performing model across all levels of imbalance. A high value of the skewed error ratio indicates strong bias and the lowest value (i.e. equal performance) is 1.
    \item \textbf{Standard deviation of performance}: For a given protected group, the standard deviation of median DSC values across all levels of imbalance was computed, i.e. $SD_{A_1} = std(\mathbf{ {D}}_{A_1})$ and $SD_{A_2} = std(\mathbf{ {D}}_{A_2})$. Again, a high value indicates strong bias, but this time the lowest value is 0.
    \item \textbf{Bias trend}: For a given protected group, the trend in bias across the different levels of imbalance was quantified by calculating a line of best fit across the five levels of imbalance. The gradient $G$ for this line is reported. A positive value of $G$ indicates an increase in performance as the representation of the first protected group increases from 0\% to 100\%.
\end{itemize}

\section{{Results}}
\label{sec:results} 

\subsubsection*{Experiment 1 - Male \vs Female:}
\begin{figure*}[!ht]
\centering
\begin{subfigure}{.45\linewidth}
    \centering
    \figuretitle{nnU-Net}
    \includegraphics[scale=0.34]{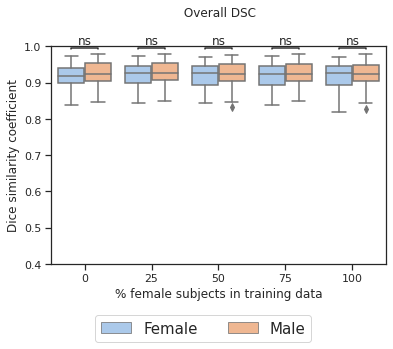}
    \caption{}\label{fig:SexDSC_nnunet}
\end{subfigure}
    \hfill
\begin{subfigure}{.45\linewidth}
    \centering
    \figuretitle{U-Net}
    \includegraphics[scale=0.34]{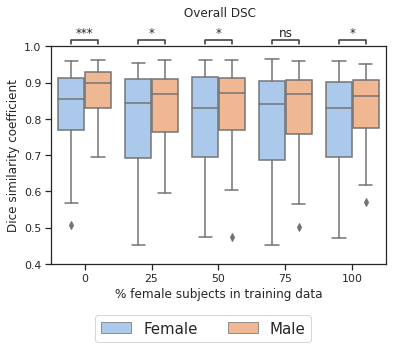}
    \caption{}\label{fig:SexDSC_unet}
\end{subfigure}

\begin{subfigure}{.45\linewidth}
    \centering
        \figuretitle{Swin-Unet}
    \includegraphics[scale=0.34]{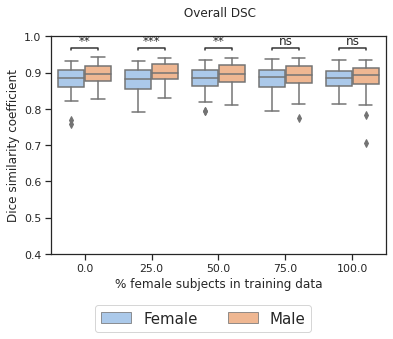}
    \caption{}\label{fig:SexDSC_swinunet}
\end{subfigure}
    \hfill
\begin{subfigure}{.45\linewidth}
    \centering
        \figuretitle{DeepLabv3+}
    \includegraphics[scale=0.34]{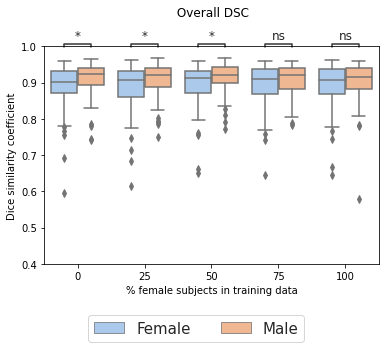}
    \caption{}\label{fig:SexDSC_deeplab}
\end{subfigure}

\caption{Overall DSC for Experiment 1 for each of the four models tested. Statistical significance was found using a Mann-Whitney U test and is denoted by **** \((p \leq 0.0001)\), *** \((0.001 < p \leq 0.0001)\), ** \((0.01 < p \leq 0.001)\), * \((0.01 < p \leq 0.05)\), ns \(( 0.05 \leq p)\).}

\label{fig:Sex_boxplots}
\end{figure*}

The results for Experiment 1 investigating the effect of imbalances in subjects' sex can be seen in \cref{fig:Sex_boxplots}. The full set of DSC scores for both of the protected groups can be found in \cref{tab: all_dsc_sex}. \cref{tab:fairness gap sex} shows the fairness gap for all of the models at each level of imbalance. For nnU-Net, there were no statistical differences in performance as the level of imbalance of the subjects changed. However, for the other three models, male subjects had statistically higher DSC scores than the females when the males were in the majority of the training set and when the training set was evenly balanced. For these models, as the proportion of female subjects in the dataset increased, accuracy parity was achieved. 

\cref{tab: sex_stats} presents the summary of bias statistics. We can see that, for the U-Net and Swin-Unet models, $PR_{male}$ was 1.4 and 1.7 times larger than $PR_{female}$ respectively. The SD of the DSC scores and $SER$ were also higher for U-Net compared to the other models and it had the largest $FG_{max}$. DeepLabv3+, which also showed an increase in performance as the proportion of females increased, had a larger $PR$, $SER$ and $SD$ for the female subjects. nnU-Net, which had accuracy parity for all proportions of female subjects in the dataset, further displayed by the smallest $FG_{max}$ and $FG_{min}$ for all of the models.

\begin{table}[!htpb]
\centering
\caption{Statistics for Experiment 1 investigating the effect of varying subjects by sex. $PR$ = performance range, $SER$ = skewed error ratio, $SD$ = standard deviation, $G$ = gradient of bias trend, $FG$ = fairness gap. The $FG$ was calculated by finding $\mathbf{{D}}_{female} - \mathbf{{D}}_{male}$. $G$ was calculated by finding the gradient of the line of best fit from 0\% females/100\% males to 100\% females/0\% males.} 
\label{tab:my-table}
\begin{tabular}{c|cc|cc|cc|cc|cc}
\toprule
\multirow{2}{*}{Model} & \multicolumn{2}{c|}{$PR$} & \multicolumn{2}{c|}{$SER$} & \multicolumn{2}{c|}{$SD$} & \multicolumn{2}{c|}{$G$} & \multicolumn{2}{c}{$FG$} \\
           & Female & Male & Female & Male  & Female & Male  & Female & Male & Max & Min   \\ \midrule
nnU-Net    &  0.0083 & 0.0020 & 1.11 & 1.03 & 0.0032 & 0.00079 & 0.00089 & $-$0.00015 & $-$0.0041 & 0.00043 \\
U-Net      & 0.026  & 0.036 & 1.18 & 1.36 & 0.011 &  0.014 & $-$0.0054 & $-$0.0072 & $-$0.042 & $-$0.024\\
Swin-Unet  &  0.0039 & 0.0068 & 1.00 & 1.07 & 0.0016 & 0.0027 & 0.000041 & $-$0.0015 & $-$0.015 & $-$0.0067\\
Deeplabv3+ &  0.012  & 0.0082 & 1.14 & 1.11 & 0.0044 & 0.0032 & 0.0016 & $-$0.0018 &$-$0.021 & $-$0.0057\\ \bottomrule
\end{tabular}
\label{tab: sex_stats}

\end{table}


\subsubsection*{Experiment 2 - Black \vs White:}

\begin{figure*}[!ht]
\centering
\begin{subfigure}{.45\linewidth}
    \centering
    \figuretitle{nnU-Net}
    \includegraphics[scale=0.34]{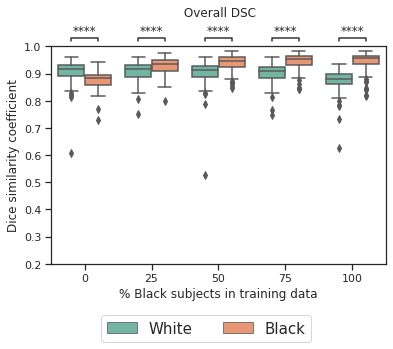}
    \caption{}\label{fig:BlackDSC_nnunet}
\end{subfigure}
    \hfill
\begin{subfigure}{.45\linewidth}
    \centering
    \figuretitle{U-Net}
    \includegraphics[scale=0.34]{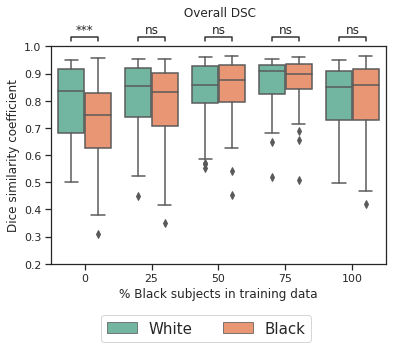}
    \caption{}\label{fig:BlackDSC_unet}
\end{subfigure}

\bigskip
\begin{subfigure}{.45\linewidth}
    \centering
        \figuretitle{Swin-Unet}
    \includegraphics[scale=0.34]{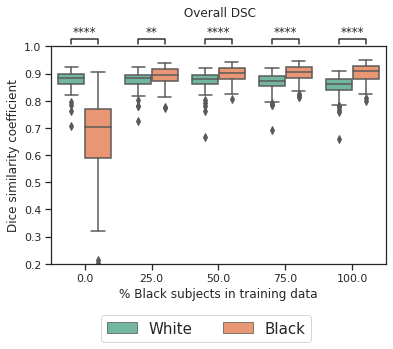}
    \caption{}\label{fig:BlackDSC_swinunet}
\end{subfigure}
    \hfill
\begin{subfigure}{.45\linewidth}
    \centering
        \figuretitle{DeepLabv3+}
    \includegraphics[scale=0.34]{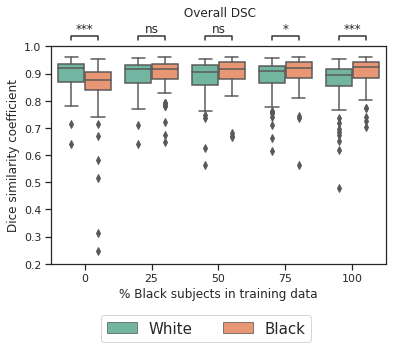}
    \caption{}\label{fig:BlackDSC_deeplab}
\end{subfigure}

\caption{Overall DSC for Experiment 2 for each of the four models tested. Statistical significance was found using a Mann-Whitney U test and is denoted by **** \((p \leq 0.0001)\), *** \((0.001 < p \leq 0.0001)\), ** \((0.01 < p \leq 0.001)\), * \((0.01 < p \leq 0.05)\), ns \(( 0.05 \leq p)\).}
\label{fig:Race_boxplots}
\end{figure*}

The results for Experiment 2 investigating the effect of imbalances in subjects' race can be seen in \cref{fig:Race_boxplots}. The full set of DSC scores for both of the protected groups can be found in \cref{tab:all_DSC_race}. \cref{tab:fairness gap race} shows the fairness gap for all of the models at each level of imbalance. As the proportion of Black subjects increased, both nnU-Net and Swin-Unet models produced significantly different DSC scores between the Black and White subjects, with the DSC scores for the Black subjects being significantly higher when their proportion of these subjects was above 25\%. However, for the DeepLabv3+ models, the performances were only significantly different at the extremes of the training dataset imbalance \ie 0\%/100\%. Lastly, for the U-Net models, only one of the performances, with 100\% White subjects and 0\% Black subjects, was significantly different.

\cref{tab: race_stats} shows the bias statistics for this experiment. For each of the models, the $PR$, $SER$ and $SD$ are higher for the Black subjects than the White subjects. Swin-Unet had the largest $PR_{Black}$, $SER_{Black}$, $SD_{Black}$ and $FG_{max}$ which can also be seen in the significantly different performances in \cref{fig:BlackDSC_swinunet}. Despite finding more significant differences in performance than the U-Net model,  DeepLabv3+ had the smallest $PR$, $SER_{Black}$, $FG_{max}$ and $FG_{min}$.


\FloatBarrier
\begin{table}[htpb]
\centering
\caption{Statistics for Experiment 2 investigating the effect of varying subjects by race.  $PR$ = performance range, $SER$ = skewed error ratio, $SD$ = standard deviation, $G$ = gradient of bias trend, $FG$ = fairness gap. The fairness gap was calculated by finding $\mathbf{{D}}_{White} - \mathbf{{D}}_{Black}$. $G$ was calculated by finding the gradient of the line of best fit from 0\% Black/100\% White subjects to 100\% Black/0\% White subjects.}
\begin{tabular}{c|cc|cc|cc|cc|cc}
\toprule
\multirow{2}{*}{Model} & \multicolumn{2}{c|}{$PR$} & \multicolumn{2}{c|}{$SER$} & \multicolumn{2}{c|}{$SD$} & \multicolumn{2}{c|}{$G$} & \multicolumn{2}{c}{$FG$}\\

 & Black & White & Black & White & Black & White  & Black &  White & Max & Min  \\ \midrule
nnU-Net  & 0.071 & 0.036 & 2.61 & 1.44 & 0.029 & 0.015 &  0.016  & $-$0.0075  & $-$0.074 & $-$0.018 \\
U-Net  & 0.15 & 0.073 & 2.52 & 1.80 & 0.059 & 0.027 & 0.029  & 0.0082 &0.088 & $-$0.0075\\
Swin-Unet & 0.21  & 0.023 & 3.25 & 1.20 & 0.090 & 0.0096 & 0.042 & $-$0.0057  & 0.18 & $-$0.014\\
Deeplabv3+ & 0.048 & 0.023 & 1.63 & 1.29 & 0.020 & 0.0088 & 0.0099 & $-$0.0053  & 0.044 & $-$0.00067 \\   \bottomrule        
\end{tabular}
\label{tab: race_stats}
\end{table}
\FloatBarrier

\clearpage 
\section{Discussion}
\label{sec:discussion}

To the best of our knowledge, this work represents the first analysis and comparison of the effect of dataset imbalance on medical imaging segmentation performance in different segmentation models. Our results show that imbalances in dataset composition affect segmentation performance for under-represented protected groups and that these biases vary for different segmentation models.

Unlike previous work \cite{Puyol-Anton2021FairnessSegmentation,Lee2022ASegmentation} which only considered nnU-Net, significant differences were found in the other three models for Experiment 1 investigating the effect of varying the proportion of subjects by sex. Despite accuracy parity being achieved for the Swin-Unet and Deeplabv3+ models where the female subjects comprised 75\% of the training set, the male subjects had a higher median DSC score at all levels of imbalance, as discussed in \cref{sec:results}.

For Experiment 2, increasing the proportion of a protected group in a training dataset increased the segmentation performance for this group, as shown in \cref{fig:Race_boxplots}, particularly \cref{fig:BlackDSC_nnunet}, \cref{fig:BlackDSC_swinunet} and \cref{fig:BlackDSC_deeplab}. However, the magnitude of this improvement varied by model. For example, $PR_{Black}$ for Swin-Unet was 0.21 but was only 0.048 for DeepLabv3+. This constitutes a difference of 9.1 times and illustrates how the same training data can produce vastly different results and biases in different models.

Although age was removed as a confounder in this work, future work should control for other confounders such as Body Mass Index or socioeconomic status. However, previous work which investigated the effect of confounders found that none of the confounders could explain the bias in nnU-Net \cite{Puyol-Anton2022FairnessSegmentation}. It is possible that these confounders, and other unknown confounders, may affect the bias found in the other models investigated in this work. In addition, only Swin-Unet, the transformer-based model, was trained using pre-trained weights whereas the other models were trained from scratch. This may have affected the convergence of the models and contributed to differences in bias noted between the four models.

Overall, for models which showed statistically significant differences in segmentation performance between protected groups, increasing the proportion of these protected groups also increased DSC scores for under-represented groups and resulted in accuracy parity. This work displays the importance of increased representation of protected groups in training datasets. More importantly, it highlights the importance of thorough model evaluation and the impact of model choice on bias. In Experiment 1, nnU-Net does not show any bias whereas U-Net does, and in Experiment 2, the reverse is true as nnU-Net shows greater bias than U-Net. This is interesting given that nnU-Net is based on the U-Net architecture but includes extra features such as connected component analysis and ensembling. We will investigate the effects of these features on bias in future work. In addition, this work highlights the effect of dataset imbalance and potential for bias in vision transformers, which will prove to be an important consideration as vision transformers are used increasingly in medical imaging applications.


\section*{Acknowledgements}
This work was supported by the Engineering \& Physical Sciences Research Council  Doctoral Training Partnership (EPSRC DTP) grant  EP/T517963/1. This research has been conducted using the UK Biobank Resource under Application Number 17806.


\printbibliography

\clearpage
\section*{Supplementary Material}
\label{sec:supplementary}

\renewcommand{\thetable}{S\arabic{table}}
\renewcommand{\thefigure}{S\arabic{figure}}

\setcounter{figure}{0}
\setcounter{table}{0}
\begin{table*}[!h]
  \caption{Training details for each of the four segmentation models. $L_{CE}$ is Cross-Entropy Loss given by $\mathcal{L}_{CE} = \sum_{i} y_i \log o_i + (1 - y_i) \log(1-o_i)$ where $o_i$ is the $i$th output of the last layer of the network and $y_i$ is the corresponding ground truth label. $L_{Dice}$ is Dice Loss given by $\mathcal{L}_{Dice} = - \frac{2 \sum_{i} o_i y_i}{\sum_{i} o_i + \sum_{i} y_i}$ \cite{Drozdzal2016TheSegmentation}.}
  \centering
  \begin{minipage}{0.5\textwidth}                          
\begin{center}
\begin{adjustbox}{center, width=\columnwidth-50pt}   
  \begin{tabular}{c|c|c|c|c|c|c}

    \toprule
Segmentation model & Number of epochs & Batch size & Pre-trained weights & Initial learning rate & Learning rate schedule & Loss function \\
\midrule
nnU-Net & 500 & 16 & No  & 0.01 & `Poly' with Nesterov momentum $=0.99$ & 0.5 $L_{CE}$ + 0.5 $L_{Dice}$ \\
Swin-Unet & 150 & 24 & Yes  & 0.01 & Momentum $=0.99$, weight decay $1e-4$ & 0.4 $L_{CE}$ + 0.6 $L_{Dice}$ \\
U-Net & 150 & 20 & No & 0.001 & Momentum $=0.99$, weight decay $1e-4$  &   $L_{CE}$ \\
DeepLabv3+ & 250 & 16 & No& 0.007 & `Poly' with power 0.9  &  $L_{CE}$\\
    \bottomrule
  \end{tabular}
  \end{adjustbox}
  \label{tab:training_params}
  \end{center}
\end{minipage}
\end{table*}

\begin{table*}[!h]
\centering
\caption{Overall median DSC for each of the four segmentation models used in Experiment 1. We report the results for each protected group and the overall test set. The train percentage signifies the percentages of protected groups used in training in the experiment. The first and second percentage values correspond with the order of the protected groups in the three experiments, \ie 0\%/100\% corresponds with 0\% female, 100\% male. Best results shown in bold.}
\resizebox{\textwidth}{!}{%
\begin{tabular}{c|ccc|ccc|ccc|ccc}
\toprule
Train & \multicolumn{3}{c|}{nnU-Net} & \multicolumn{3}{c|}{U-Net} & \multicolumn{3}{c|}{Swin-Unet} & \multicolumn{3}{c}{DeepLabv3+} \\
Percentage & Female & Male & All & Female & Male & All & Female & Male & All & Female & Male & All \\ \midrule
0\%/100\% & 0.920 & 0.924 & 0.923 & \textbf{0.856} & \textbf{0.898} & \textbf{0.869} & 0.886 & 0.897 & 0.891 & 0.902 & \textbf{0.922} & 0.912 \\
25\%/75\% & \textbf{0.928} & \textbf{0.925} & \textbf{0.927} & 0.843 & 0.867 & 0.859 & 0.884 & \textbf{0.899} & 0.890 & 0.908 & 0.921 & 0.913 \\
50\%/50\% & 0.927 & 0.924 & 0.926 & 0.831 & 0.872 & 0.858 & 0.884 & 0.897 & 0.892 & \textbf{0.914} & 0.921 & \textbf{0.918} \\
75\%/25\% & 0.925 & 0.924 & 0.925 & 0.842 & 0.867 & 0.858 & \textbf{0.887} & 0.894 & \textbf{0.892} & 0.910 & 0.920 & 0.916 \\
100\%/0\% & 0.925 & 0.923 & 0.924 & 0.830 & 0.862 & 0.856 & 0.884 & 0.892 & 0.889 & 0.909 & 0.914 & 0.913 \\ \bottomrule
\end{tabular}%
}
\label{tab: all_dsc_sex}
\end{table*}
\begin{table*}[!h]
\centering
\caption{Overall median DSC for each of the four segmentation models used in Experiment 2. We report the results for each protected group and the overall test set. The train percentage signifies the percentages of protected groups used in training in the experiment. The first and second percentage values correspond with the order of the protected groups in the experiment, \ie 0\%/100\% corresponds with 0\% Black, 100\% White. Best results shown in bold.}
\resizebox{\textwidth}{!}{%
\begin{tabular}{c|ccc|ccc|ccc|ccc}
\toprule
Train & \multicolumn{3}{c|}{nnU-Net} & \multicolumn{3}{c|}{U-Net} & \multicolumn{3}{c|}{Swin-Unet} & \multicolumn{3}{c}{DeepLabv3+} \\
Percentage & White & Black & All & White & Black & All & White & Black & All & White & Black & All \\ \midrule
0\%/100\% & 0.916 & 0.884 & 0.894 & 0.835 & 0.747 & 0.780 & \textbf{0.884} & 0.703 & 0.847 & \textbf{0.919} & 0.875 & 0.894 \\
25\%/75\% & \textbf{0.918} & 0.936 & 0.925 & 0.856 & 0.834 & 0.840 & 0.882 & 0.896 & 0.888 & 0.915 & 0.916 & \textbf{0.916} \\
50\%/50\% & 0.913 & 0.946 & 0.925 & 0.859 & 0.878 & 0.865 & 0.880 & 0.901 & 0.889 & 0.907 & 0.916 & 0.912 \\
75\%/25\% & 0.910 & 0.953 & \textbf{0.926} & \textbf{0.908} & \textbf{0.900} & \textbf{0.903} & 0.871 & 0.907 & \textbf{0.890} & 0.908 & 0.919 & 0.911 \\
100\%/0\% & 0.882 & \textbf{0.956} & 0.905 & 0.850 & 0.857 & 0.854 & 0.861 & \textbf{0.909} & 0.881 & 0.896 & \textbf{0.923} & 0.908 \\ \bottomrule
\end{tabular}%
}
\label{tab:all_DSC_race}

\end{table*}
\begin{table}[]
\centering
\caption{Fairness gaps $FG$ for Experiment 1 investigating the effect of varying subjects by sex. The fairness gap was calculated by finding $\mathbf{{D}}_{female} - \mathbf{{D}}_{male}$. The train percentage signifies the percentages of protected groups used in training in the experiment. The first and second percentage values correspond with the order of the protected groups in the experiment, \ie 0\%/100\% corresponds with 0\% female, 100\% male.}
\label{tab:fairness gap sex}
\begin{tabular}{c|c|c|c|c}
\toprule
Training percentage & nn-Unet & U-Net   & Swin-Unet & DeepLabv3+ \\
\midrule
0\%/100\%           & $-$0.0041 & $-$0.042 & $-$0.011  & $-$0.021     \\
25\%/75\%           & 0.0025  & $-$0.024 & $-$0.015   & $-$0.013    \\
50\%/50\%           & 0.0030  & $-$0.041 & $-$0.013  & $-$0.0070     \\
75\%/25\%           & 0.00043  & $-$0.025 & $-$0.0067   & $-$0.010    \\
100\%/0\%           & 0.0021  & $-$0.032 & $-$0.0082   & $-$0.0057   \\
\bottomrule
\end{tabular}
\end{table}

\begin{table}[]
\centering
\caption{Fairness gaps $FG$ for Experiment 2 investigating the effect of varying subjects by race. The fairness gap was calculated by finding $\mathbf{{D}}_{White} - \mathbf{{D}}_{Black}$. The train percentage signifies the percentages of protected groups used in training in the experiment. The first and second percentage values correspond with the order of the protected groups in the experiment, \ie 0\%/100\% corresponds with 0\% Black, 100\% White.}
\label{tab:fairness gap race}
\begin{tabular}{c|c|c|c|c}
\toprule
Training percentage & nn-Unet & U-Net   & Swin-Unet & DeepLabv3+ \\
\midrule
0\%/100\%           & 0.031 & 0.088 & 0.18   & 0.044     \\
25\%/75\%           & $-$0.018  & 0.022 & $-$0.014   & $-$0.00067    \\
50\%/50\%           & $-$0.033  & $-$0.019 & $-$0.021   & $-$0.0084     \\
75\%/25\%           & $-$0.043  & 0.0083 & $-$0.036   & $-$0.011    \\
100\%/0\%           & $-$0.074  & $-$0.0075 & $-$0.048   & $-$0.027   \\
\bottomrule
\end{tabular}
\end{table}

\end{document}